# RNAMunin: A Deep Machine Learning Model for Non-coding RNA Discovery

Lauren M. Lui*[†1] and Torben N. Nielsen*[†2]

[1]Environmental Genomics and Systems Biology Division, Lawrence Berkeley National Laboratory, Berkeley, CA, USA.

[2]Jorgmundir, Seattle, WA, USA.

*Correspondence: lmlui@lbl.gov, torben@jorgmundir.life

[†]Contributed equally to this work.  Order is alphabetical.

## Abstract

Functional annotation of microbial genomes is often biased toward protein-coding genes, leaving a vast, unexplored landscape of non-coding RNAs (ncRNAs) that are critical for regulating bacterial and archaeal physiology, stress response and metabolism. Identifying ncRNAs directly from genomic sequence is a paramount challenge in bioinformatics and biology, essential for understanding the complete regulatory potential of an organism. This paper presents RNAMunin, a machine learning (ML) model that is capable of finding ncRNAs using genomic sequence alone. It is also computationally viable for large sequence datasets such as long read metagenomic assemblies with contigs totaling multiple Gbp. RNAMunin is trained on Rfam sequences extracted from approximately 60 Gbp of long read metagenomes from 16 San Francisco Estuary samples. We know of no other model that can detect ncRNAs based solely on genomic sequence at this scale. Since RNAMunin only requires genomic sequence as input, we do not need for an ncRNA to be transcribed to find it, *i.e.*, we do not need transcriptomics data. We wrote this manuscript in a narrative style in order to best convey how RNAMunin was developed and how it works in detail. Unlike almost all current ML models, at approximately 1M parameters, RNAMunin is very small and very fast.

## Introduction

For decades, the annotation of microbial genomes has been dominated by the identification of protein coding genes, creating a picture of compact genomes optimized for metabolic efficiency. However, this protein-centric view overlooks a vast and functionally critical component of the microbial regulatory network: sequences that are transcribed but not translated. These molecules often function by virtue of their ability to fold into constructs that participate in controlling gene expression at both transcriptional and translational levels, helping microbes to rapidly adapt to changing environments, manage stress, and orchestrate complex behaviors[1,2]. These molecules are commonly referred to as non-coding RNAs (ncRNAs), *i.e.*, they do not code for proteins. We will use the term ncRNA *sensu lato* to refer to any sequence of RNA that shares significant features matching known Rfams[3] as determined by the software suite Infernal[4].

The microbial ncRNA world is remarkably diverse. The most well known ncRNAs include ribosomal RNAs (rRNAs) and transfer RNAs (tRNAs). However, ncRNAs also encompass small RNAs (sRNAs) which typically range in size from 20 - 200 nucleotides and act by base pairing with target mRNAs to modulate their translation or stability. Structured ncRNAs include riboswitches[5], which are cis-regulatory elements typically within mRNA 5'-UTR regions that directly bind to specific metabolites to control the expression of downstream genes, acting as autonomous genetic switches. Some ncRNAs have even been found to hide inside messenger RNAs (mRNAs)[6] and conversely, sequences that code for small proteins have been found inside ncRNAs[7]. The discovery of these types of ncRNAs has revealed a hidden layer of regulation that is integral to almost all aspects of microbial life. There are many other named groups of RNAs, but in this work, we will not make the distinction between different classes of ncRNAs.

Despite their importance, the systematic identification of ncRNA genes remains a major bioinformatic challenge. Unlike protein coding genes which are defined by long, conserved open reading frames (ORFs) and predictable codon usage patterns[8], ncRNAs lack universal, easily identifiable sequence features. This is because secondary structure, such as hairpins, frequently defines their function. Non-coding RNA sequences are often poorly conserved across diverse species, while their functional secondary structures can be difficult to predict accurately from genomic sequence alone. There have been significant efforts at discovering ncRNAs in microbial isolates[9] and in an environmental microbiome[10] but these either rely on transcriptomic sequencing (where some ncRNAs may not be detected because they are only transcribed under specific conditions) or are too labor intensive to scale to very large datasets such as the ones arising from long read metagenomic sequencing of environmental samples[11].

The challenge for ncRNA detection is to develop computational methods that are fast enough to scan large datasets such as metagenomes but also reliably distinguish the faint signals of a true ncRNA gene from the vast background of possibly nonfunctional intergenic DNA or unannotated ORFs. The computational methods must be practical, meaning they must be able to run in reasonable time on generally available hardware. Currently, the state-of-the-art (SOA) for finding ncRNAs in genomes or metagenomes is using covariance models[4], which are probabilistic profiles of sequence and secondary structure of RNA families. However, these are specific to each RNA family. To illustrate how impractical it is to search for noncoding RNAs in large datasets, we provide an example using a bacterial genome.

Throughout this work, we will use a 3.75 Mbp *Curtobacterium* genome we sequenced and assembled as an example. It is part of a project to generate the most accurate complete bacterial genomes possible from long read Nanopore data. We extracted the DNA, sequenced it, assembled it and annotated it; *i.e.*, we have all of the information we need to track down any possible annotation anomalies related to sample handling and bioinformatics processing. For the purposes of this work, we will refer to it as LT001001. We used Pyrodigal to annotate the protein coding genes[12]. Using a single thread, these annotations were completed in 1.81 s. We also used Infernal 1.1 to annotate the same genome against Rfam 15.0[3]. The Rfam annotations required 1,140 s with an average of 16.76 threads. Ignoring the difference in threads used, the ratio of the time to annotate protein coding genes vs non-coding RNA genes is 1:630. Thus, we can extrapolate that Infernal would take a little over 35 days to annotate a 10 Gbp metagenome against the 4178 families in the current version of Rfam[3]. Finding the protein coding genes for this example metagenome would be done in a little over 80 minutes. There are certainly ways to speed up the annotations of both the protein coding genes and the ncRNAs, but no matter how we optimize the execution of current SOA bioinformatics methods - really just Infernal - annotating ncRNAs this way is impractical for modern long read metagenomes. As Rfam continues to grow, the situation will only worsen. Ironically, we expect growth in Rfam to be - at least in part - due to the diversity of ncRNA exposed in the long read metagenomes and metatranscriptomes.

Our interest is not in predicting RNA structure, rather we are focused on using machine learning to find RNA sequences that appear likely to have structure. Given such sequences, there are many applications that can help find sequences with matching structure and refine them into full covariance models such as those in Rfam; RNAlien from the Vienna RNA Package is one such[13,14] and GLASSgo [15] is another. Deep learning methods to predict RNA secondary and tertiary structure have made strides in recent years[16–18]. Reducing the search space for functional noncoding RNAs will massively speed up the search of noncoding RNAs in large sequence datasets.

The machine learning (ML) model we present in this work has been trained on a large corpus of known representatives of Rfam models and it is capable of recognizing all such representatives we have presented it with. We call this model RNAMunin after Munin from Norse mythology. Munin is one of the ravens of Odin the Allfather, who flies across the world to gather information and represents memory. We think that this is an appropriate name as the model represents the memory of RNA in data that was gathered from multiple metagenomes. RNAMunin has flagged many sequences that are not represented by Rfam and all such that we have looked at fold into credible secondary structures when run through RNAfold[19]. Upon manual inspection, all that we have seen have typical ncRNA structures such as bulges, stem loops and hairpins[20] and thus we infer that these are likely real ncRNA. Quoting Doug Adams[21], "If it looks like a duck, and quacks like a duck, we have at least to consider the possibility that we have a small aquatic bird of the family Anatidae on our hands."

# Model

RNAMunin is an ML model that takes genomic sequences as input - we assume in FASTA format - and predicts ncRNAs that function by virtue of their secondary structure potential. The ultimate goal is to take a sequence of nucleotides and determine what ncRNAs it contains and where they are located; *i.e.*, approximate beginning and end along the input sequence.

A classic ML approach would be to take known ncRNA sequences, pad them to a common length and attempt to train a model to predict if a given sequence is an ncRNA or not. A key difficulty of this approach is that ncRNAs come in many different lengths (one reason for the thousands of families in Rfam); 50-3000 nucleotides (nt) are common with the shorter ones being the most abundant. Examples of common short ncRNA (< 100nt) are riboswitches and also tRNAs, the most common known ncRNA sequence we are aware of. Common long ncRNA in prokaryotic genomes are the small subunit rRNA, which is typically ~1,500 nt in length and the large subunit rRNA, which is almost 3,000 nt. The amount of padding required to accommodate the range of ncRNA sequence lengths is too high. This approach is also described as global; *i.e.*, the model tries to match the entire ncRNA from end to end.

We instead decided on a local approach. Specifically, for each nucleotide in a sequence, we ask if that nucleotide appears to be part of an ncRNA or not. To make the decision, we look at a fixed number - 50 is what we settled on - of nucleotides upstream or downstream of the nucleotide in question. Anthropomorphizing, we asked each nucleotide in the sequence a simple question: based on the 50 nucleotides that are your neighbors on either side, do you think you are part of an ncRNA?

This leads to a simple binary classification problem. For each nucleotide, the feature vector consists of the 50 nt upstream and downstream in addition to the nucleotide itself (Figure 1). The target is set to 1 if the midpoint of the sequence - position 51 using 1-based coordinates - is contained in an ncRNA and to 0 otherwise. Using 101 for the length of each feature vector is somewhat arbitrary. The choice was made to keep the model small enough to run quickly even on very limited hardware while also allowing a significant amount of secondary RNA structure to be covered. A length of 101 is sufficient for a whole tRNA sequence and many riboswitches. Thus, although we picked a somewhat arbitrary length for the feature vector, we will demonstrate that this model works remarkably well.

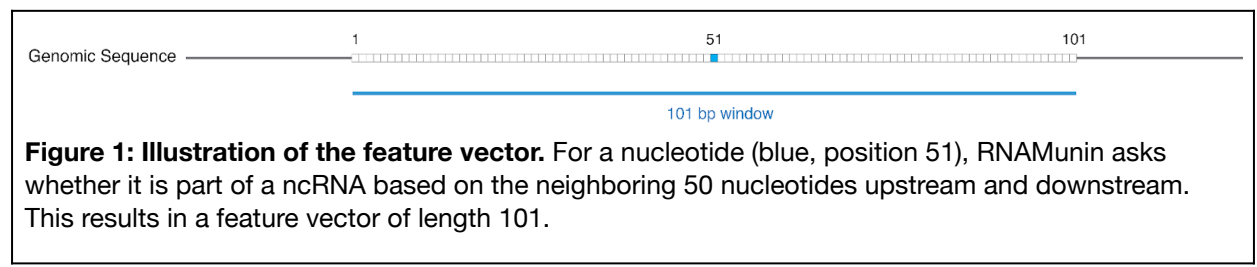

**Figure 1: Illustration of the feature vector.** For a nucleotide (blue, position 51), RNAMunin asks whether it is part of a ncRNA based on the neighboring 50 nucleotides upstream and downstream. This results in a feature vector of length 101.

The following code block shows the actual model we used for training. It is a simple Keras sequential model[22]. The first layer specifies the shape of the input. As outlined in the section on training data, the input consists of one-hot encoded sequences of length 101. There are three 1-D convolutional layers. All use a kernel size of 4 with no padding. All convolutional layers use 'relu' as their activation function. The number of filters goes from 128 in the first convolutional layer through 256 in the second and finally to 512 in the third. There are two max-pooling layers; the first uses a pool size of 4 and the second uses a pool size of 3. Following the last max-pooling layer all the filters are flattened resulting in 2,048 nodes. These are then passed into a dense network with 192 nodes and a 'relu' activation function. Following that layer we inserted a dropout layer with a parameter of 0.3 to guard against overfitting. The final layer has a single node and uses 'sigmoid' for an activation function. The output should be regarded as the probability that the node in position 51 on the training sequence is in an ncRNA.

```
model = keras.Sequential([
    keras.Input(shape = (101, 4)),
    layers.Conv1D(filters = 128, kernel_size = 4, activation = 'relu'),
    layers.MaxPooling1D(pool_size = 4),
    layers.Conv1D(filters = 256, kernel_size = 4, activation = 'relu'),
    layers.MaxPooling1D(pool_size = 3),
    layers.Conv1D(filters = 512, kernel_size = 4, activation = 'relu'),
    layers.Flatten(),
    layers.Dense(units = 192, activation = 'relu'),
    layers.Dropout(0.3),
    layers.Dense(units = 1, activation = 'sigmoid')]
)
```

We use binary cross entropy for loss and the Adam optimizer. It is important to note that the decisions made by the model for two different nucleotides are completely independent.

# Data

To train a ML model for a binary classification problem, both positive and negative training sequences are needed. To generate training data, we need known ncRNAs. We considered publicly available datasets as a source for the ncRNAs we needed. Obtaining known high quality data is often difficult[23]. There is a multitude of advertised collections of such ncRNAs, but there is little information regarding the curation and quality. Fortunately for us - while we are computational - we are also able to generate our own data. We have collections both of isolates and large metagenomes where we know the exact provenance of the data since we generated it ourselves from DNA extraction through sequencing, basecalling, assembly and annotation. To ensure high quality, we only include sequences found by `cmsearch` (Infernal) that are non-truncated matches to Rfam families.

## San Francisco Estuary Metagenomes: Collection of High-Quality Data

In the summer of 2022 and the winter of 2023, we obtained water samples from the San Francisco Estuary (SFE) courtesy of the United States Geological Service (USGS). The samples were collected along a path that the USGS traverses on a monthly basis for water quality monitoring[24]. We received 8 summer samples and 8 winter samples. We extracted DNA from all the samples and sequenced each sample using two Oxford Nanopore Technologies (ONT) PromethION flow cells. We then basecalled and assembled the data to produce 16 long read metagenomes. Full details on the materials and methods used are in Lui and Nielsen 2024[11].

In total, from the SFE metagenomes there were 59,971,032,290 nucleotides from 2,559,532 contigs. We used Infernal and Rfam to find all of the ncRNA corresponding to Rfam families in the datasets. It should be noted that we used Rfam 12.0 for this task[25]. These searches resulted in 174,198,538 nucleotides of ncRNA on 1,070,073 sequences (Figure 2).

Note that only 0.29% of the total nucleotides were identified by Infernal as belonging to a known Rfam. We used an in-house script based on seqtk [26] to extract the sequences and reverse complement as needed to get all of the sequences from the correct strand.

There was significant overlap among the sequences and we elected to remove duplicates. For this, we used seqkit with parameters "rmdup -s -i -P" [27]. This resulted in 89,591,837 nucleotides across 362,620 unique sequences. While this deduplicated set has no duplicated sequences, a significant number of the sequences are representatives of just a few Rfams (Figure 2).

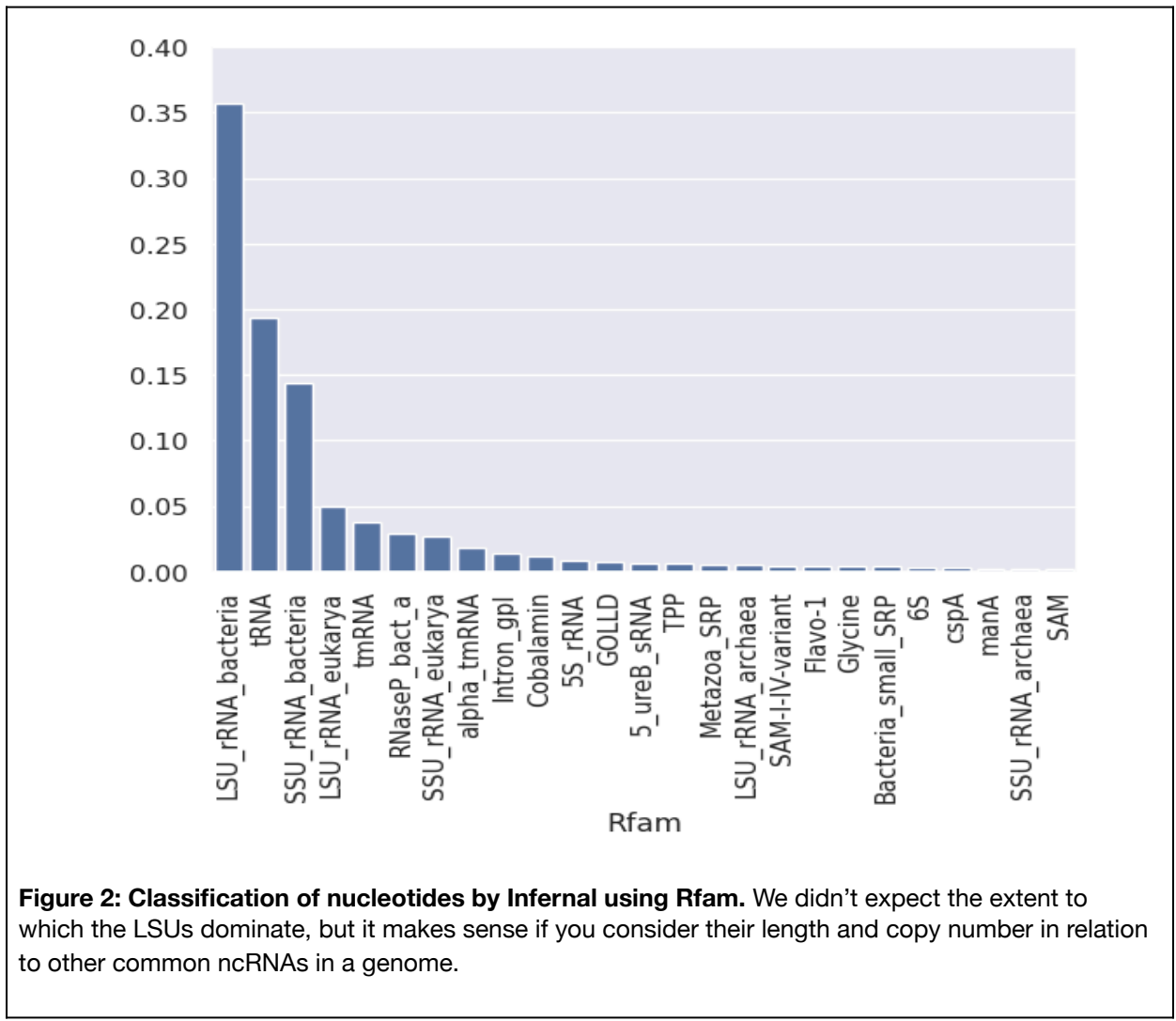

**Figure 2: Classification of nucleotides by Infernal using Rfam.** We didn't expect the extent to which the LSUs dominate, but it makes sense if you consider their length and copy number in relation to other common ncRNAs in a genome.

## Generation of the ncRNA training set

To generate the training sequences, we started with a full length ncRNA sequence containing all upper case nucleotides; *i.e.*, A, T, G and C. We then calculated the distribution of the nucleotides in the sequence and generated a sequence of nucleotides using lower case; *i.e.*, a, t, g and c, of four times the length of the original sequence randomly drawn from this distribution. We prepended half of the random sequence and appended the other half to the ncRNA. We then extracted all subsequences of length 101 from the resulting nucleotide sequence and set the target to 1 if position 51 was a nucleotide written in upper case and to 0 otherwise. The use of upper and lower case is merely for ease of processing since it lets us tell if a specific nucleotide came from a known ncRNA sequence or was part of the randomly generated sequence.

The prepended and appended randomized sequences provide the negative samples for our training set while the central portion is an ncRNA matching a known Rfam. We intentionally generated the randomized portion using the nucleotide distribution from the known ncRNA. We wanted to prevent the model from using simple distributional statistics to distinguish. This has been known to affect performance of machine learning models[28]. We are well aware that the negative samples are not going to be representative of what our model will see at test time. The genomic sequences surrounding ncRNAs are generally not random. But because we want to be able to find novel sequences as well as known ones, we did not want to simply use the genomic sequences surrounding the ncRNAs on the contigs where they are found as negative feature vectors. They might contain novel sequences which we wish to find and using them as negative samples would bias the model against them.

We used this deduplicated set of ncRNA sequences as the core of the dataset building process outlined above. At the very end, we one-hot encoded the nucleotide sequences so that A/a: (1, 0, 0, 0), T/t: (0, 1, 0, 0), G/g:(0, 0, 1, 0) and C/c:(0, 0, 0, 1). This results in a tensor of dimension (101, 4).

As indicated above, known structural ncRNAs do not account for a very large proportion of assembled sequence data. Based on our data, it is safe to say that such ncRNAs are rare. While we believe there are a significant number of ncRNAs we have yet to find, we do not think that ncRNAs exhibiting secondary structure account for more than an order of magnitude greater than what we are already finding using Infernal against Rfam. Because structural ncRNAs are rare, we need to be concerned about the size of the negative sample set as well as its composition which we previously discussed. The choice of four times as much negative as positive data was a result of experimentation.

# Training

We used our baseline system (described in the testing section below) for training. It took approximately three weeks to train for 150 epochs. This is a significant amount of time, but it is also not something that needs to be repeated. The model results are excellent. The figure below shows the epoch accuracy and epoch loss for both training and validation data (Figure 3). The difference in training and validation accuracy is approximately 0.5% and we do not consider it significant. There is no evidence that the model is memorizing the data.

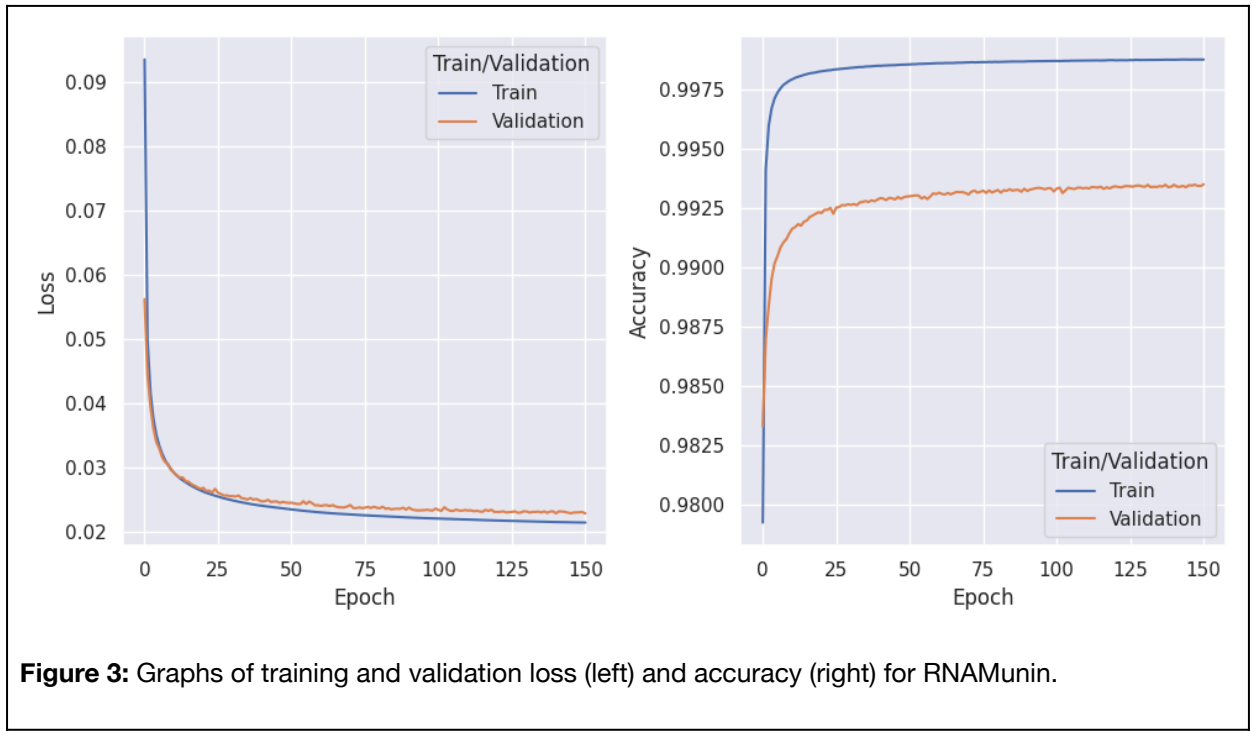

**Figure 3:** Graphs of training and validation loss (left) and accuracy (right) for RNAMunin.

# Testing

For testing, we used the LT001001 *Curtobacterium* genome. We are very concerned with being able to annotate large metagenomes, but here we wanted to focus on correctness and for that, an isolate genome is preferable. Also, a goal of long read metagenomics is to fully resolve the community; *i.e.*, decompose it into its constituent genomes in which case everything reduces to being able to correctly annotate isolates. If the model works for isolate size genomes, we see the rest as a matter of scaling.

As previously described, the model is formulated as a binary classification problem. That is, for each nucleotide in the sequence we pass it, RNAMunin tries to decide if that nucleotide is part of an ncRNA based on the neighborhood it is in, which itself and the 50 nucleotides on each side (Figure 1). This score is commonly interpreted as a probability and based on a threshold, RNAMunin decides "yes" if the score is greater than or equal to the threshold and "no" otherwise.

Running a complete genome of length N through the model yields two length N sequences of real numbers between 0 and 1 as a result. The first sequence represents the scores assigned to nucleotides on the forward strand and the second is the scores assigned to nucleotides on the negative strand. By themselves, these sequences of scores are difficult to use. We usually convert both of them into runs; *i.e.*, lists of consecutive positions of scores that all exceed a fixed threshold. A threshold of 0.5 is common for binary classification problems, but for this model, we chose a threshold of 0.9. The choice of 0.9 is somewhat arbitrary. We made it to reduce the number of runs found and in practice, it performs well for our testing. That isn't to say that lower thresholds cannot produce worthwhile results. The classic problem with ML models is that they are not generally explainable.

Thus, the outcome of testing LT001001 is two sets of ranges of positions - one for the forward strand and one for the reverse strand - where the consecutive scores are all greater than or equal to 0.9. Each of these ranges represents a putative ncRNA. We reduced the number we need to look at by filtering any runs shorter than 20 nt. That comes to 1,937 putative ncRNA sequences produced by the model. In describing our test results, we will focus on four points:

1. Does the model find all of the ncRNAs found by Infernal against Rfam? That is a basic requirement.
2. Does the model identify any novel ncRNAs that are not found by Infernal against Rfam?
3. How does the model fare against a random input? That is, how does it perform if we give it a random sequence with the same GC content and length as LT001001 as input?
4. Will the model be able to produce results with reasonable resources? That is, can we process a substantial long read metagenome (*e.g.*,10 Gbp) in a week or less?

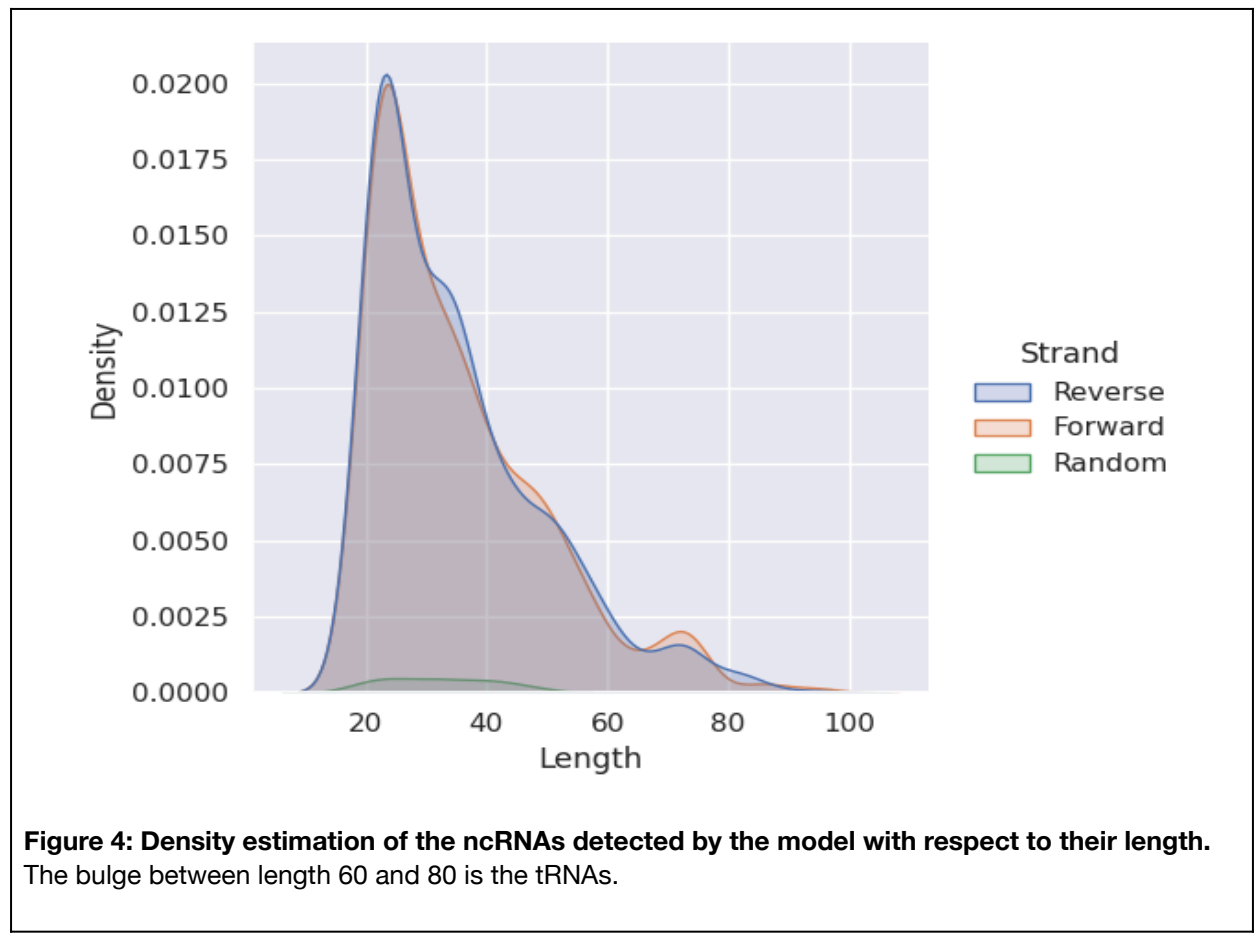

**Figure 4: Density estimation of the ncRNAs detected by the model with respect to their length.** The bulge between length 60 and 80 is the tRNAs.

## Does RNAMunin find all of the ncRNAs found by Infernal against Rfam?

RNAMunin finds everything that Infernal against Rfam does. We verified this by going through the GFF file we generated from Infernal against Rfam 15.0. For each entry in the GFF file, we checked which strand - forward or reverse - it was on and then we checked if it was matched by one of the ranges output by the model for the relevant strand. tRNAs were commonly matched with a difference of +/-1 for the beginning and end. Matches to the LSU_rRNA_bacteria family were matched by 6 ranges separated by a few nucleotides each. This may be due to the fact that LSUs are typically divided into domains with sections and there may not be enough secondary structure for the model to pick up. It should be noted that almost all of the scores in the matching ranges were 1.0. That is, if RNAMunin indicated any confidence that the nucleotide it was looking at could be part of an ncRNA, it was generally very high.

## Does RNAMunin identify any novel ncRNAs that are not found by Infernal against Rfam?

RNAMunin finds many novel putative ncRNA sequences. There are 66 Infernal identified Rfam sequences. But the model - counting both strands - produces 1,937 putative ncRNA sequences. That leaves 1,871 putative novel ncRNA sequences in the LT001001 genome.

## How does RNAMunin fare against random input?

We generated a random sequence of nucleotides with length and GC matching LT001001 and we ran the model on it. The two strands had a total of 27 runs with scores all greater than or equal to 0.9 and of length at least 20 nucleotides. Out of the 27, 7 were at least 40 nucleotides long. Figure 4 shows one of length 42 folded with RNAfold[14]. The MFE is -22.48 kcal/mol or -0.54 kcal/mol/nt which is not unusual for real ncRNAs[6].

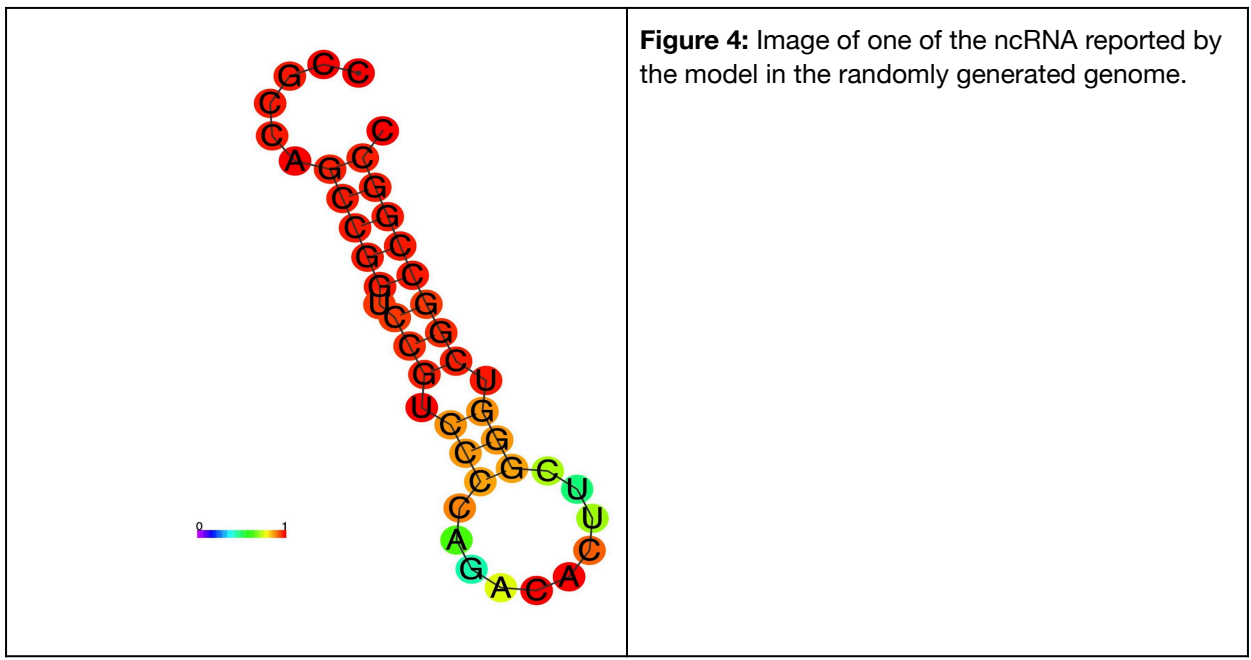

**Figure 4:** Image of one of the ncRNA reported by the model in the randomly generated genome.

Since input is random, we do not want to label this a false positive. Random sequences of nucleotides can have secondary structure that looks like ncRNAs. The model is trained on sequences that are known to have secondary structure and it is to be expected that it finds some in random nucleotide sequences. What is worth noting is that the putative ncRNAs from the randomly generated sequence is only 1.39% of the number found in the LT001001 genome.

### Will RNAMunin be able to produce results with reasonable resources?

All development and benchmarking is done on a system with an AMD Ryzen Threadripper PRO 7975WX 32-core CPU with 512 GB of main memory and 8 TB of NVMe SSD. We also have an NVIDIA RTX 6000 Ada Generation GPU. That is our baseline system. Running the model with LT001001 as input takes 94 seconds (wall clock time) using 5 cores and less than 5 GB of main memory. The test script is straight Python with no attempt at optimization. From monitoring runs, we believe that most of the runtime is spent to prepare and download the input to the GPU.

Using the length of the genome, we can again extrapolate how long it would take to run the model on a larger dataset. The performance metrics described above are for a genome that is 3,748,858 bp in length. The 10 Gbp long read metagenome that we wish to be able to process is 2,668X as long and - since execution time is directly proportional to length - we expect it to take 69 hours; *i.e.*, a little less than 3 days. To put this number in perspective, we note that it is approximately the same amount of time, perhaps slightly less, it takes to base call the raw long read data that becomes a 10 Gbp metagenome on the same hardware. In the context of the kind of data we focus on, we consider this to be reasonable.

For isolates and smaller metagenomes, we see no reason it should not run on a modern Apple laptop with Apple silicon. All of our development was done using the Keras API and it can run using Apple Metal.

# Results and Discussion

We developed and trained an ML model - RNAMunin - for the detection of stretches of nucleotide sequence that have features in common with known ncRNAs. RNAMunin was trained on a large corpus of known ncRNAs from long read metagenomes we generated from the San Francisco Estuary[11]. The corpus was constructed using Infernal to annotate all of the sequences matching known Rfams. Presented with a genomic sequence as input, the model attaches to each nucleotide on each strand - forward and reverse - a probability that the nucleotide is part of an ncRNA. The probabilities are purely local judgments; the model does not keep any knowledge of probabilities assigned to nucleotides upstream or downstream.

RNAMunin is fast. It runs in a little over 1.5 minutes for the test genome (3.75 Mbp) on our hardware. Much of the speed comes from the size. Technically, RNAMunin is a Deep Neural Network (DNN), but it is small, weighing in at only a little over 1 million parameters. RNAMunin is developed to focus on a specific task and to do it well. This is the core of the original Unix Philosophy; do one thing, do it well and expect the output of your program to become the input for several other programs[29].

RNAMunin is also accurate. It achieves accuracies in excess of 99% on both the training and the validation data. Moreover, every member of Rfam found by Infernal is consistently assigned a probability of 0.9 or greater over almost the entire sequence produced by Infernal. Any variances tend to be +/- 1 nucleotides at the ends except for the very long sequences such as those from LSUs which have short breaks between what appears to be separate domains; *i.e.*, these short breaks are likely linkers.

For all of the LT001001 genome, RNAMunin flags runs - all longer than 20 nt - with each nucleotide having probability 0.9 or greater totaling 81,395 bp. This is across both strands and it comes to 2.15% of the total nucleotides in LT001001 classified as ncRNA. Infernal using Rfam annotates only 19,378 nucleotides across both strands, which is only 0.52%. RNAMunin suggests that almost 4X as much sequence as is annotated via Rfam are likely to be ncRNA. In terms of number of sequences, RNAMunin generates 1,937 versus the 66 produced by Infernal against Rfam; *i.e.*, a factor close to 30X. We believe the long microbial ncRNAs have been found and what we are looking at is many shorter sequences.

We intend to mine the long read metagenomes we have for potential ncRNAs. Metagenomes provide diversity that has not been captured by laboratory cultures. We know that RNAMunin is capable of flagging sequences that have features similar to known ncRNAs. Moreover, we know that it is practical to run it on very large long read metagenomes. This will allow us to extract likely candidates and use tools like RNAlien or GLASSgo to produce covariance models. Similar to the many novel findings of species and protein coding genes in metagenomes in the last 15 years, we expect to find new ncRNAs.

# Acknowledgements

The *Curtobacterium* isolate was provided courtesy of Professor Jennifer Martiny at the University of California Irvine from a collection maintained by her lab.

We would like to thank USGS for collecting samples for us, especially Erica Nejad and the crew of USGS R/V David H. Peterson. We cannot overemphasize the value of the support we have received for this project from USGS.

This work was supported by the Laboratory Directed Research and Development Program of Lawrence Berkeley National Laboratory under U.S. Department of Energy (DOE) Contract No. DE-AC02-05CH11231.

# References

1. Ponath, F., Hör, J. & Vogel, J. An overview of gene regulation in bacteria by small RNAs derived from mRNA 3′ ends. *FEMS Microbiol Rev* **46**, fuac017 (2022).

2. Morfeldt, E., Taylor, D., von Gabain, A. & Arvidson, S. Activation of alpha-toxin translation

in Staphylococcus aureus by the trans-encoded antisense RNA, RNAIII. *The EMBO Journal* **14**, 4569–4577 (1995).

3. Ontiveros-Palacios, N. *et al.* Rfam 15: RNA families database in 2025. *Nucleic Acids Res* **53**, D258–D267 (2025).

4. Nawrocki, E. P. & Eddy, S. R. Infernal 1.1: 100-fold faster RNA homology searches. *Bioinformatics* **29**, 2933–2935 (2013).

5. Narunsky, A. *et al.* The discovery of novel noncoding RNAs in 50 bacterial genomes. *Nucleic Acids Res* **52**, 5152–5165 (2024).

6. Dar, D. & Sorek, R. Bacterial noncoding RNAs excised from within protein-coding transcripts. *MBio* **9**, (2018).

7. Gray, T., Storz, G. & Papenfort, K. Small Proteins; Big Questions. *J Bacteriol* **204**, e0034121 (2022).

8. Hyatt, D. *et al.* Prodigal: prokaryotic gene recognition and translation initiation site identification. *BMC Bioinformatics* **11**, 119 (2010).

9. Stav, S. *et al.* Genome-wide discovery of structured noncoding RNAs in bacteria. *BMC Microbiol* **19**, 66 (2019).

10. Gelsinger, D. R. *et al.* Regulatory Noncoding Small RNAs Are Diverse and Abundant in an Extremophilic Microbial Community. *mSystems* **5**, (2020).

11. Lui, L. & Nielsen, T. Decomposing a San Francisco estuary microbiome using long-read metagenomics reveals species- and strain-level dominance from picoeukaryotes to viruses. *mSystems* (2024) doi:10.1128/msystems.00242-24.

12. Larralde, M. Pyrodigal: Python bindings and interface to Prodigal, an efficient method for gene prediction in prokaryotes. *Journal of Open Source Software* **7**, 4296 (2022).

13. Eggenhofer, F., Hofacker, I. L. & Höner Zu Siederdissen, C. RNAlien - Unsupervised RNA family model construction. *Nucleic Acids Res* **44**, 8433–8441 (2016).

14. Lorenz, R. *et al.* ViennaRNA Package 2.0. *Algorithms for Molecular Biology* **6**, 1–14 (2011).

15. Lott, S. C. *et al.* GLASSgo - automated and reliable detection of sRNA homologs from a single input sequence. *Front. Genet.* **9**, 124 (2018).

16. Predicting RNA structure and dynamics with deep learning and solution scattering. *Biophysical Journal* **124**, 549–564 (2025).

17. Singh, J., Hanson, J., Paliwal, K. & Zhou, Y. RNA secondary structure prediction using an ensemble of two-dimensional deep neural networks and transfer learning. *Nature Communications* **10**, 1–13 (2019).

18. Lang, M., Litfin, T., Chen, K., Zhan, J. & Zhou, Y. Benchmarking the methods for predicting base pairs in RNA–RNA interactions. *Bioinformatics* **41**, btaf289 (2025).

19. Gruber, A. R., Lorenz, R., Bernhart, S. H., Neuböck, R. & Hofacker, I. L. The Vienna RNA websuite. *Nucleic Acids Res.* **36**, W70–4 (2008).

20. Durbin, R., Eddy, S. R., Krogh, A. & Mitchison, G. *Biological Sequence Analysis*. (Cambridge University Press, Cambridge, England, 2007).

21. Adams, D. *Dirk Gently Detect*. (Simon & Schuster, 1987).

22. Keras: Deep Learning for humans. https://keras.io.

23. Hwang, H., Jeon, H., Yeo, N. & Baek, D. Big data and deep learning for RNA biology. *Exp Mol Med* **56**, 1293–1321 (2024).

24. Cloern, J. E. Patterns, pace, and processes of water-quality variability in a long-studied estuary. *Limnol. Oceanogr.* **64**, (2019).

25. Nawrocki, E. P. *et al.* Rfam 12.0: updates to the RNA families database. *Nucleic Acids Res*

**43**, D130–D137 (2014).

26. *[No Title]*. (Github).

27. Shen, W., Le, S., Li, Y. & Hu, F. SeqKit: A Cross-Platform and Ultrafast Toolkit for FASTA/Q File Manipulation. *PLOS ONE* **11**, e0163962 (2016).

28. González, M., Durán, R. E., Seeger, M., Araya, M. & Jara, N. Negative dataset selection impacts machine learning-based predictors for multiple bacterial species promoters. *Bioinformatics* **41**, (2025).

29. Salus, P. H. *A Quarter Century of UNIX*. (Addison-Wesley Professional, 1994).

30. Faure, G., Ogurtsov, A. Y., Shabalina, S. A. & Koonin, E. V. Role of mRNA structure in the control of protein folding. *Nucleic Acids Res.* **44**, 10898–10911 (2016).